\documentclass[11pt]{article}

\usepackage[dvips]{epsfig}
\usepackage{a4wide}

\newcommand{\beq}{\begin{equation}}
\newcommand{\eeq}{\end{equation}}
\newcommand{\bea}{\begin{eqnarray}}
\newcommand{\eea}{\end{eqnarray}}
\newcommand{\bean}{\begin{eqnarray*}}
\newcommand{\eean}{\end{eqnarray*}}
\newcommand{\lab}[1]{\label{#1}}
\newlength{\smallwidth}
\newlength{\red}
\newcommand{\deldel}{{\del^2 \! } }
\newcommand{\e}{\mbox{e}}
\newcommand{\s}{\sigma}

\newcommand{\ta}{\tau}
\newcommand{\rh}{\rho}

\newcommand{\ep}{\epsilon}

\newcommand{\om}{\omega}
\newcommand{\la}{\lambda}
\newcommand{\G}{\Gamma}
\newcommand{\Om}{\Omega}
\newcommand{\Si}{\Sigma}
\renewcommand{\a}{\alpha}
\newcommand{\g}{\gamma}
\renewcommand{\b}{\beta}
\newcommand{\del}{\delta}

\newcommand{\ax}{{\bar x}}
\newcommand{\ap}{{\bar p}}

\newcommand{\ih}{{\tilde h}}

\newcommand{\cT}{{\cal T}}
\newcommand{\cA}{{\cal A}}

\newcommand{\cG}{{\cal G}}
\newcommand{\cE}{{\cal E}}
\newcommand{\cR}{{\cal R}}
\newcommand{\cO}{{\cal O}}
\newcommand{\cF}{{\cal F}}

\newcommand{\cX}{{\cal X}}

\newcommand{\cU}{{\cal U}}
\newcommand{\cV}{{\cal V}}
\newcommand{\bra}{\left<}
\newcommand{\ket}{\right>}

\newcommand{\nn}{\nonumber}
\newcommand{\1}{\mbox{1 \hspace*{ -1em} 1  }}

\newcommand{\9}{\partial}
\newcommand{\de}{\delta}

\newcommand{\tr}{\mbox{tr}}

\newlength{\oldarraycolsep}

\newlength{\abswidth}
\begin{document}

\thispagestyle{empty}

\rightline{\today}
\rightline{HD-THEP-98-32}
\rightline{hep-th/9807145}

\begin{center} 

\vspace*{\fill}

{\bf{Dissipative Time Evolution of Observables in Non-equilibrium 
Statistical Quantum Systems}  }

\bigskip
  
Herbert Nachbagauer\footnote{nachbaga@thphys.uni-heidelberg.de}  \\

\medskip 
{\em 
Institut f\"ur Theoretische Physik, Universit\"at Heidelberg, 
Philosophenweg 16, D-69120~Heidelberg, Germany}  

\bigskip \bigskip

{\begin{minipage}{\abswidth}

{\begin{center} {\bf Abstract }  \end{center} }

We discuss differential-- versus  
integral--equation based methods describing out--of thermal 
equilibrium systems and emphasize the importance of a well 
defined reduction to statistical observables. Applying the 
projection operator approach,
we investigate on the time evolution of expectation values of linear
and quadratic polynomials in position and momentum for a 
statistical anharmonic oscillator with quartic potential. 
Based on the exact integro-differential equations of motion, we 
study the first and naive second order approximation which 
breaks down at secular time-scales. A method is proposed to 
improve the expansion by a non--perturbative resummation of 
all quadratic operator correlators consistent with energy 
conservation for all times.  Motion cannot be described by 
an effective Hamiltonian local in time reflecting non-unitarity 
of the dissipative entropy generating evolution. We 
numerically integrate the consistently improved equations 
of motion for large times. We relate entropy to the uncertainty 
product, both being expressible in terms of the observables 
under consideration.

\medskip 


\end{minipage} }

\end{center} 

\vspace*{\fill}

\newpage 

\section{Introduction} 

Non-equilibrium aspects of quantum field systems are of 
actual interest.  Among other problems,
one challenge is the description of time evolution of macroscopic
quantities such as the expectation value of the field strength in systems out of 
thermal equilibrium. The problem, essentially the basic question for 
macroscopic dynamics of statistical systems, and 
thus interesting in itself, also plays a prominent role in context with 
cosmological inflationary phases.  Although a plethora of 
equivalent descriptions of the collisionless limit are 
available, it is difficult to elaborate methods which allow to go beyond this 
thermodynamically trivial limit. One purpose of this paper is to promote 
a strategy based on which time evolution can be described
in a consistent and systematic way.    

Although the basic concepts of thermodynamic statistical theory
are known for a long time, there still appears to be 
some confusion about terminology and underlying concepts 
of non-equilibrium systems and their dynamical description. Another task 
of this work is to introduce, review and discuss some of the basic
ideas in a brief but systematic way. Technically, we adopt the projection 
operator method as one consistent method to describe time evolution 
of physical quantities in statistical systems.   

To keep complexity minimal, we consider the toy model 
of the quantum Duffing oscillator for practical calculations, which 
nevertheless is a system of sufficient complexity to exhibit important 
features of non-equilibrium thermodynamics. For systematic 
reasons, we discuss some properties of that system in the 
statistically trivial, entropy conserving, approximation first. 
We present criteria for permissible initial data, study static 
solutions and their stability, construct the first order 
integrals of motion and the first order effective 
Hamiltonian and Lagrangian. The system exhibits phenomena of 
parametric resonance which best may be displayed in the small 
coupling limit. 

Finally, we attack our main task and investigate on proper non-equilibrium 
features. In a first step, that requires the study of second order 
effects which, however, turn out to violate energy conservation at long
time scales in the strict power--series expansion.  
The main result of this investigation is to improve the second order 
result by taking into account the non--unitary effective evolution of 
observables which renders the approximation scheme self--consistent
and solves the problem of non--conservation of constants of motion. 
In the appendix, we explicitly derive the basic uncertainty--entropy 
relation for the set of quadratic observables.

\section{Definitions and general properties}

(i) A mixed state of a quantum system (configuration), both in zero
dimensional quantum mechanics as well as in quantum field theory, 
is described by a density operator, which, in order to allow a 
probability interpretation, must be a hermitian trace-class operator with 
positive eigenvalues. It is an intrinsic feature of quantum 
systems that the density matrix is fictive in the sense that only 
its diagonal elements correspond to physical probabilities, the 
other dependencies being phases which enter in expectation
values via interference effects\footnote{That is the fundamental 
difference to classical phase space averages.}.  Alternatively, 
one may characterize a configuration 
by the expectation values of hermitian operators. A generic 
set of those representing a complete set of observables is 
given by the mutually orthogonal hermitian projectors constructed 
from the eigenvectors of the density matrix. A complete set of 
(not necessarily commuting) observables contains all information 
about the density matrix such that any observable can be expressed 
in terms of the complete set. 

(ii) The system may be assigned a dynamical structure. Motion is 
defined as a sequence of possible states having a constant 
expectation value of the Hamiltonian operator, the energy of the 
configuration. Quantum mechanical time parameterizes those 
configurations which are assumed to have time--independent 
probabilities and phases for autonomous systems. The time evolution 
generated by the Hamiltonian can be expressed by first order differential 
operator--equations in time for the density matrix  (von~Neumann 
equation) in the Schr\"odinger representation. Hermitian Hamilton
operators generate unitary time evolution which in
turn is necessary for compatibility of the probability interpretation
with time evolution. Once the problem of time evolution of the 
mixed state is solved for a given initial configuration, observables can be 
calculated as expectation values from the evolved density matrix. 

(iii) The statistical description of a system is based on a reduction procedure 
selecting an in general much smaller number of (macroscopic) 
observed quantities out of the complete set of observables. This subset defines 
the level of description. The process of reduction from the density operator to 
the set of the so--called relevant quantities in general involves loss of 
information. Here, we will concentrate on levels of descriptions that
once chosen, will be kept fixed during time evolution. That constraint
can be relaxed too if necessary \cite{balian86}.

(iv) The reduction can be dynamically trivial if it commutes with time 
evolution. In that case of dynamical closure, the Liouvillian maps the operators of the 
level of description on a linear combination of them. The observables of 
a level of description at a certain time are sufficient to determine them at 
any time and evolution induces neither information gain nor information 
loss at the level of description, the associated entropy being constant.
On account of the canonical commutation relation, the most general 
Hamiltonian admitting a finite dimensional dynamically closed level of 
observation can contain constant, linear and quadratic expressions in 
position and momentum operators\footnote{Regardless of additional 
dependencies on other c-number quantities, including time and classical 
field strength, we call it free (effective) Hamiltonian.}. The corresponding 
dynamically closed sets of observables correspond to sums of polynomials of 
finite order in the canonical operators.  
If the reduction is non-trivial, one can extend the level of observation
to render it trivial. That may involve an infinite number of operators in 
which case the system is a truly interacting one, and the 
only dynamically closed set of operators corresponds to the complete 
set of observables.  

(v) For truly interacting dynamical systems, the complete initial 
density matrix influences observables at later times. Its definition
calls for an additional principle to construct it from the reduced set of 
initial data. Information theory proposes to apply Shannon's   
theory of entropy \cite{shan49} to that physical problem. Jaynes' principle 
of maximum entropy \cite{jayn89} fixes the generalized canonical 
density operator as initial condition. It can be shown not to contain more 
information than the initial set of observables. We want to point out that 
this choice is the statistically most probable, but the underlying concept of 
ensemble averages of identical systems evolving from variant
initial preparations does not imply the actual preparation of the system 
in that state.  

We confront two strategies to arrive at a description of the motion of observables. 

(A) Solve the complete problem of time evolution 
for the mixed initial state and extract the interesting quantities. 
The reduction procedure explicitly is postponed after time evolution.

(B) Construct equations of motion for the relevant quantities.
They turn out to be integro-differential equation in time necessary to 
include the effects of the past history on the relevant observables.

Solution (A) appears to be be more attractive for field theorists
since methods of field theory can be applied with rather little 
modification. The problem boils down to solve the initial value 
problem for mixed states for which the appropriate tool 
of the closed--time--path method has been developed \cite{chou85}.   
We sketch the systematic procedure (A) according to the program
outlined in (i)-(v), and discuss associated technical and conceptual 
problems.  

Starting with the equations of motion in differential form, 
one gets an infinite hierarchy of first order differential
equations for the complete set of observables, corresponding
to the Schwinger--Dyson hierarchy of equal time correlators,
and analogous to the Bogoliubov-Born-Green-Kirkwood-Yvon 
hierarchy of many--particle physics \cite{bale75}. The hierarchy 
may be represented in compact form in terms of a classical
effective potential generating the equal--time correlators 
\cite{effpo} by transforming to the Wigner representation. 
In principle, the reduction amounts to eliminate the irrelevant quantities 
resulting in a finite set of infinite order differential equations for the
reduced set of observables the integration of which involves an infinite 
number of initial conditions. At that point the statistical assumption 
about the initial density matrix enters to determine all derivatives 
of the observables at the initial time $t_0$. The essential feature of 
those effective equations of motion is non--locality in time which 
becomes apparent when converting them into integral equations.
Suppose we only knew the relevant observables at a later time $t_1>t_0$, 
then neither retrodictions nor predictions can be made on account of
the lacking knowledge on the derivatives at $t_1$, 
or the complete history of observables. In that sense, reduction 
necessarily goes along with current temporal information loss. A 
complete calculation of the present observables can only be done on 
grounds of the complete past initial density matrix where missing 
information was induced by the principle of maximum
uncertainty. The corresponding reduced entropy at later times is 
always larger than the entropy of the initial state which introduces 
thermodynamic irreversibility in a natural way.

We want argue that, strictly speaking, the cut hierarchy is not a permissible 
dynamical approximation. There, one keeps only those
equations which involve the time derivatives of the variables of 
interest. In order to arrive at a closed set of equations, the 
irrelevant quantities generally still being present in the system
have to be expressed somehow in terms of the relevant ones.
There is no systematics of how to do that. In some particular
cases, as the large N--expansion of $O(N)$ symmetric scalar theories,
the hierarchy by chance closes without further assumptions 
\cite{largeN}. In the classical theory of gases Boltzmann introduced 
the famous Sto{\ss}zahlansatz for two particle correlators at that point.   
Generally dynamics then is represented by a finite closed set of 
effective (macroscopic) equations of motion. Of course no
statistical assumptions about the complete initial state has to be 
inferred.  

We want to point out that the truncation of the differential 
hierarchy explicitly alters the dynamics of the 
system in an uncontrolled way. Any finite 
order hierarchy of equations of motion exhibits features of a
closed dynamic system discussed in (iv) and is thus 
a priori inequivalent to the reduced truly interacting evolution. 
It is always local in time which presumes memory loss at 
microscopic scales and automatically reduces to a Markovian 
description, even in cases where the complete 
system behaves differently. Some truncations are compatible with 
the existence of a macroscopic effective Hamiltonian 
(Lagrangian) being a function of a finite set of effective variables. 
The very existence of that Hamiltonian description, however, 
automatically conserves the relevant reduced von~Neumann entropy 
such that the system can neither exhibit equilibration nor thermodynamic 
irreversibility in the strict sense\footnote{Effective irreversibility,
sometimes misleadingly related to ergodicity \cite{bale75}, may occur in 
infinite--dimensional systems \cite{coop96}, but has to be justified 
a posteriori and cannot be concluded on grounds of an approximation
scheme.}. Also, due to the Liouville--theorem the volume of cells 
in the phase space remains constant under effective Hamiltonian evolution. 
The volume, on the other hand, corresponds to the classical particle 
number density, which enters in the entropy functional of Boltzmann. 

The argument may best be illustrated 
for the class of effectively Gaussian approximations which
can be obtained by time dependent Hartree--Fock variational methods 
\cite{TDHF}, the  method of equal--time Green functions \cite{coop96,boya98}, 
optimized expansions \cite{okop} or the truncation
of the Schwinger--Dyson hierarchy to the one and two 
particle Green function. All those methods are physically equivalent
in the Gaussian limit where the Hamiltonian is assumed to have 
free form, or, equivalently the density matrix is approximated by 
a generalized Gaussian wave package. One considers time dependent spatially 
bilocal functions --- essentially equal--time two point correlators ---
as coefficients in the quadratic expressions which are determined 
by the corresponding variational and expansion 
methods. But those bilocal coefficients enter in the generalized 
uncertainty which is a constant of time in the Gaussian approximation. 
On account of the uncertainty--entropy relation, also entropy remains constant 
contradicting information loss inherent in the incomplete description.
Under certain conditions, the effective Gaussian theory can even be 
shown to have a representation in terms of a free theory after 
requantisation \cite{bena97}. Although this approximation 
may supply a fairly good description at short terms or close 
to equilibrium, they are thermodynamically trivial approximations and 
cannot account for full long term evolution since they effectively 
correspond to the collisionless\footnote{Strictly speaking, the term 
collisionless is not precise since the limit does account for effective 
elastic scattering.} limit of the physical system. We will show that in 
fact an effective Gaussian density matrix, whatever its time dependence 
be, does not evolve to the Gaussian approximation of the interacting 
density matrix at large times.
The 'equilibration' observed in those approximations \cite{coop96,wett97} 
can thus not be identified with physical entropy generating processes, 
but converges to static values by dephasing effects.

In general, the dynamics of a cut hierarchy effectively corresponds
to current preparation of the system to have the canonical 
density matrix of the level of observation. 
Thus, dynamical closure is attained at the price of effectively time
dependent probabilities being typical for non--autonomous
systems. The procedure presumes that the density matrix decoheres 
for macroscopic observables including the existence of a classical 
limit for the variables of the level of observation. Such an effective theory 
may exist, depending on the particular system under consideration, but it
remains to show that it represents motion also at macroscopically large time
scales.

The problems with the differential representations can be avoided 
if we adopt solution (B) which intrinsically accounts for 
non--locality in time. The practical method we apply is the 
projection operator method in Schr\"odinger representation
which yields the Robertson equation \cite{robi66} being
equivalent to the generalized Langevin--equation \cite{balian86,proj}
in the Heisenberg representation. Although those
equations of motion are closed in the variables of the 
level of observation only, expectation values 
of arbitrary operators can be expressed by non--local 
time dependence. The exact reduced equations of motion turn out to have a
rather complicated structure and suitable approximation schemes 
have to be developed within that framework. Splitting off a dynamically
closed part from the Hamiltonian, the entropy generating contributions 
involve quadratic and higher powers in the remaining 
truly interacting Hamiltonian. There, a non-unitary 
operator enters, responsible for time evolution in the 
projected subspace. Approximating that operator by a unitary one  
again amounts to introduce a Gaussian approximation which 
differs formally only at third order from the exact solution but 
nevertheless involves unphysical consequences. 
It conflicts with entropy generated at second order and,
more severely, results in non-conservation of the energy 
at secular scales. 

\section{Basic Setup}
We briefly sketch the result of the projection operator approach 
\cite{balian86,proj}. 
For convenience, we define a level of observation by the finite set of
hermitian operators $\cE=\lbrace \cF_\nu \rbrace $ including $\cF_0=\1$. 
For the corresponding operator expectation values 
$g_\nu(t)= \tr \left( \cF_\nu  \rho(t) \right) $, 
the closed exact equation of motion is found to read 
\beq
 \frac{d}{dt}  \tr \left( \cF_\nu \cR(t) \right)  
= -i \, \tr \left( \cF_\nu  L \circ \cR(t) \right) 
  - \int_{0}^t \! dt' \, \tr \left( \cF_\nu L \circ T(t,t') \circ Q(t')
\circ L \circ \cR(t') \right) .  \lab{final}
\eeq
The accompanying canonical density operator
$\cR(t) = \exp ( - \mu_\nu (t) \cF_\nu )$ which minimizes the 
entropy in $\cE$ contains time dependent
Lagrange multipliers $\mu(t) $ which are functions 
of $g_\nu(t)$ such that $g_\nu(t) = \tr \left( \cF_\nu \cR(t) \right)$. 
We emphasize that $\cR(t)$ does not evolve with $L$ and must not be 
confused with the density-operator $\rho(t)$ of the system. 
All trace expressions 
at the r.h.s.\ of (\ref{final}) can at least in principle be expressed in terms 
of the $g_\nu$, and the system is a closed integro-differential
equation in the c-number expectation values.  
The projector $Q(t)$ is defined by 
\beq
\tr ( \cO_1 Q(t) \circ \cO_2 ) = \tr  ( \cO_1  \cO_2 )  -  
 \frac{\9 \tr ( \cO_1 \cR(t) )  }{\9 g_\nu(t) }  \,  \tr ( \cF_\nu \cO_2 )   ,
\eeq
and the non-unitary evolution operator $T(t,t')$ is a solution of 
\beq
\frac{\9}{\9 t} T(t,t') = - i Q(t) \circ L \circ T(t,t') \quad \mbox{and} \quad  
\frac{\9}{\9 t'} T(t,t') = i  T(t,t') \circ  Q(t') \circ L \lab{Tgl} 
\eeq 
with  initial condition $T(t,t) = 1 $.

Here, we will complete $\cE$ such that 
the action of the Hamiltonian $[H,\cX]= L \circ \cX $ can  be split 
into $L=L_0+L_1$, and the free part be dynamically closed 
with respect to the level of observation, 
$L_0 \circ \cF_\nu  = [ H_0 , \cF_\nu ] = \Om_{\nu\mu} \cF_\mu$.
In that case, in the integral (\ref{final}), the complete Liouvillian can be replaced by
$L_1$.  

Expectation values of operators 
which are not in the linear hull of $\cE$ get additional contributions to their 
$\cR$-averages,
\beq
 \tr (\cO \rho(t) ) = \tr \left( \cO \cR(t) \right) 
- i \int_0^t dt' \tr \left( \cO T(t,t') \circ Q(t') \circ L \circ  \cR(t') \right) 
\eeq
where again the dynamically closed part in $L$ does not contribute to the 
integral. The integral vanishes for $\cO \in \cE$. 

\newpage 

\section{Zero-dimensional system} 

We will investigate on the time evolution of an anharmonic oscillator 
defined by the Hamiltonian
\beq 
H = H_0 + H_1, \quad H_0 = \frac{1}{2} (p^2 + s x^2) , \quad 
H_1 = \frac{\la}{2} x^4 
\eeq
with canonical commutator $[x,p] = i$. The parameter $s=\pm 1$
allows to switch to the  broken symmetry case.  We define the 
level of description by a dynamically closed 
set of observables for $H_0$ given by $\cE= \lbrace \1,
x,p,x^2,p^2,x p + p x\rbrace  $. 
That set plays a preferred role since, due to its bilinear nature, 
the accompanying density operator $\cR(t)$ is
quasi-Gaussian which allows to apply a modified Wick theorem in the 
evaluation of expectation values. Inclusion of the quadratic 
variables also ensures the existence of the accompanying 
canonical operator, and allows for direct comparison with 
approximations of Gaussian type. The equations of 
motion are found to read 
\bea
\frac{d}{dt} \ax  &=& \ap \lab{eom1} \\ 
\frac{d}{dt} \ap &=& -s  \ax + 4 \la \ax^3  - 6 \la \ax x_2  + \Si_{p}(t)  \lab{eom2}\\
\frac{d}{dt} x_2 &=& w \lab{eom3}  \\
\frac{d}{dt} p_2 &=&  8 \la \ap \ax^3 - s  w - 6 \la w x_2 + \Si_{p^2}(t)\lab{eom4} \\
\frac{d}{dt} w   &=& 2 p_2 + 8\la \ax^4 - 2 s x_2-12 \la x_2^2+\Si_{w}(t)\lab{eom5}
\eea
with $\ax=\langle x \rangle ,\ap=\langle p\rangle ,x_2 = \langle x^2\rangle ,
p_2 =\langle p^2\rangle ,w=\langle xp +px\rangle $ and
\beq
\Si_{\nu}(t) = - \int_0^t dt' \tr ( \cF_\nu L_1\circ T(t,t') \circ Q(t')\circ L_1 \circ 
\cR(t') ) \lab{lls} 
\eeq
accounting for entropy increase. 
Note that $\dot \ax$ and $\dot x_2$ do not get further corrections since the 
potential is a function of position only. The energy is of the form
\bea
\tr \left( H \rho(t)\right)  &= &\frac{1}{2} ( p_2 + s  x_2 ) + \frac{\la}{2} ( 3 x_2^2 - 2 x^4 ) + 
\ep(t) , \nn \\ 
\ep(t) &=&  - i \int_0^t dt' \tr ( H_1 T(t,t')\circ Q(t') \circ L_1 \circ \cR(t') ) ,\lab{eint} 
\eea
and is conserved in time. 
The key challenge is to find a  sensible approximation for the operator $T(t,t')$. 
However it is worthwhile to study the first order system without integrals 
first. 
 
\section{First order approximation}

\subsection{Permissible initial conditions}
 
The first order system may be integrated numerically with initial conditions
for $\ax,\ldots , w$. Those as well as their time evolved counterparts are 
subject to the positivity conditions, $x_2 \geq 0,p_2 \geq 0$,
$\del^2 x =  x_2 - \ax^2 \geq   0 , \, 
\del^2 p = p_2 - \ap ^2  \geq   0 $,
and the uncertainty relation $(\del^2 w = w - 2 \ax \ap )$ 
\beq
\cA^2 := (\del^2 x) (\del^2 p) - \frac{1}{4} (\del^2 w )^2 \geq  \frac{1}{4}. 
\eeq
which follows immediately from Schwarz' inequality relation\footnote{
This condition generally gives stronger constraints than the usual procedure 
to minimize the positive--definite expression $\bra ( A - \la B)^2 \ket \geq 0 .$ }  
$\bra A ^2 \ket \bra  B^2\ket  \geq \vert \bra A B \ket \vert ^2 .$
Coefficients that do not fulfill that relations cannot appear as expectation
values averaged with a physical accompanying density operator $\cR$.

\subsection{Static solutions and their stability}
Static solutions of (\ref{eom1}--\ref{eom5}) are found to read 
\beq
\ax  =\pm \sqrt{ \frac{s + 6 \la x_{2}}{4 \la } } ,\quad 
p_2  = - \frac{1}{4 \la}  ( s + 2 \la x_{2} ) ( s + 6 \la x_{2} ) , \quad 
w=\ap = 0 .
\eeq 
In the unbroken case $s=1$, for any choice of $x_2$  the value 
$p_2$ is always negative.  We conclude that whatever motion be, 
for permissible initial conditions the system never 
attains a state where all observables become time independent. 

In the broken case $s=-1$, however, the  positivity requirement for $ p_2 $
permits static solutions in the 
range $ 1/(6 \la) \leq x_{2}  \leq 1/(2 \la) $. They may have 
expectation values for position with $| \ax  | \leq  1/\sqrt{2 \la } $, 
where the bounds correspond to the classical local minima of the potential.  
Classically, one would not expect static  solutions for values of $\ax$ other 
than the minimum $1/\sqrt{2 \la }$. 
An additional restriction is given by the uncertainty relation, 
\beq
( 1 - 2 \la  x_{2} )^2 ( -1 + 6 \la x_{2} )  \geq 4 \la^2 .
\eeq
For $\la \ll 1 $, this condition  narrows the region of static solutions 
$ 1/(6 \la) + 3 \la /2 + \cO( \la^2) \leq x_{2}  \leq 1/(2 \la) - 1/\sqrt{2} + \cO(\la) $
by a small amount. For increasing $\la$,  the window gets smaller  and finally 
completely vanishes at $\la = \frac{2}{9} \sqrt{\frac{2}{3}} 
\approx 0.18 $, corresponding to $x_{2} = \frac{5}{4} \sqrt{\frac{3}{2}} \approx 1.53 $ and 
$\ax = (27/2)^{1/4}/2  \approx 0.96$. For larger 
couplings, no static solutions exist. In that case, the minimal quantum 
uncertainty suffices to overcome the potential barrier and the motion
can escape the initial half. 

To study the stability of the static points in the broken phase, 
we consider a set of initial conditions
with $\ap=w=0$ and $x_2, p_2$ satisfying the staticity requirement. We chose
$\ax(0) = \ax^{\mbox{static}} + \de $, where $\de /\ax \ll 1$ to model perturbed 
initial conditions. For $\de > 0$
a numerical integration shows that 
the position observable periodically increases to a maximum, at which the width 
$\deldel x$ develops a local minimum. The motion of $\ax$ remains in the initial half, 
but the amplitude is large with respect to the initial coordinate.

For   $\de < 0$, the qualitative picture is completely different. There the average
position performs an oscillation to the negative initial value, 
while the width oscillates only slightly.  
It is surprising that motion is just opposite to the classically expected roll--down 
in the potential. Again quantum effects largely dominate over the 
classical picture. 

In both cases, the motion of the position coordinate does not remain near the 
initial preparation. In that sense, static solutions are not attractive,
and cannot be regarded as stable.  

\subsection{First order integrals of motion}

Rewriting the system (\ref{eom1}--\ref{eom5})
it in terms of the quantum widths $X = \sqrt{\deldel x},\, 
P=\sqrt{\deldel p},\, W^2 = \deldel w  $,
\beq
\frac{d}{dt} W^2 =2 P^2  - 2  (s  + 6 \la x_2 ) X^2  ,\quad
\frac{d}{dt} P^2  =  - (s  + 6 \la x_2 )  W^2  ,\quad
\frac{d}{dt} X^2 =   W^2 , 
\eeq
one verifies the time independence of $\cA$. 
It is remarkable that the uncertainty appears as constant of motion which 
moreover in independent of the coupling. Physically, on account of the 
uncertainty--entropy  relation (see Appendix), constant 
uncertainty means that quantum effects do not increase entropy. To
first order, the value of the set of observables $\cE$ at a given time 
is sufficient to determine their time evolution locally in time, and to express all 
observables of the system. 

A second constant of motion is given by
\beq
h  = \frac{1}{2} p_2 + \frac{1}{2} s  x_2 +\frac{\la}{2}(3 x_2^2- 2 \ax^4)  \lab{h}
\eeq
which  equals the expectation value of the Hamiltonian averaged with $\cR$. 
That statement is not
so trivial as seems at first sight.  It is a consequence of the fact of dynamical 
closure of $H_0$ which makes the corrections in the integral in  (\ref{eint})
of second order.  

Let us further make a comparison with the standard perturbative approximation.
There, one expands $H$ around the classical expectation value 
of momentum and position, i.e. 
$x = \ax + \de x$, where $\de x $ is supposed to be small. 
Applying that approximation to the Hamiltonian and keeping only 
quadratic terms in $\de x$, one finds for the $\la $--dependent 
contribution to the energy $\la ( 6 \ax^2 x_2 - 5 \ax^4)/2 $, which differs from
$h$ if $\de x$ is not small, to wit, if quantum or statistical corrections 
become important which is in fact the case. 
Note also that in the case of symmetric configurations $\ax=0$, the quartic potential part
does not contribute to energy in that approximation at all. Thus
the standard perturbative first order effective potential does not coincide 
with the consistent Gaussian dynamical potential, in agreement with \cite{coop96}.

\subsection{First Order Effective Hamiltonian and effective Lagrangian} 

The energy integral can be used to eliminate $\ap, p_2,w$ from
the system (\ref{eom1}--\ref{eom5}),
\bea
\ddot \ax &=& - s  \ax + 4 \la \ax^3 - 6 \la \ax x_2 \lab{29} ,\\   
\ddot x_2 & = & 4 h - 4 s  x_2 - 18 \la x_2^2 + 12 \la \ax^4 \lab{30} .
\eea
This system is not a Hamiltonian system in the kinetic momenta
$\dot \ax, \dot x_2$. A necessary condition
of the existence of a Hamiltonian  $\dot \ax^2 + \dot x_2^2 +
J(\ax , x_2 ) $ would require $J$ to be a solution of 
\beq
\frac{\9 J}{\9 \ax} = - \ddot \ax ( \ax,x_2), \quad  
\frac{\9 J}{\9 x_2} = - \ddot x_2 (\ax,x_2) .
\eeq 
However, the integrability  condition $\9^2 J/(\9\ax \9 x_2) =
\9^2 J/(\9 x_2 \9\ax ) $ is not fulfilled by the corresponding 
expressions of (\ref{29},\ref{30}). 

Nevertheless, an effective Hamiltonian which reproduces the 
equations motion of (\ref{eom1}-\ref{eom5}) together with the constant 
of uncertainty $\cA^2$ can be constructed. We introduce the 
canonical pairs $(\ax,\pi)$ and $(x_2,\pi_2)$ and cast the 
conserved energy (\ref{h}) in canonical variables. Then,
$p_2$ becomes 
a function of $(\ax,x_2,\pi,\pi_2)$  and has to be consistent with the 
Hamiltonian equations of motions, 
\beq
\ap = \dot \ax =   \frac{\9 H}{\9 \pi}  =  \frac{1}{2} \frac{\9 p_2 }{\9 \pi} , \qquad
w = \dot x_2 = \frac{\9 H}{\9 \pi_2} = \frac{1}{2} \frac{\9 p_2 }{\9 \pi_2} . \lab{33a}
\eeq  
On the other hand, the constant uncertainty requires $p_2$ to be a solution 
of the partial differential equation
\beq
\frac{1}{16} \left(  \frac{\9 p_2 }{ \9 \pi_2 }  \right)^2  - 
\frac{\ax}{4}  \frac{\9 p_2}{ \9 \pi_2 } \frac{\9 p_2}{\9 \pi }
+ \frac{x_2}{4} \left(   \frac{\9 p_2 }{ \9 \pi }  \right)^2 = p_2 X^2   - \cA^2
\eeq
Splitting off a summand $\cA^2 / X^2$ the remaining function has to be 
a homogeneous expression in $\pi,\pi_2$ of order two. We find 
\beq 
p_2 = \frac{\cA^2}{X^2} + ( 4 x_2 \pi_2^2 + 4 \ax \pi_2 \pi + \pi^2 ) , \quad 
\ap = ( 2 \ax \pi_2 + \pi ),
\eeq
which relates observables to the  canonical variables. 
The corresponding Hamiltonian function has the form
\bea
H(\ax,\pi; x_2 , \pi_2) &= & 2 x_2 \pi_2^2 + 2 \ax \pi_2 \pi + \frac{1}{2} \pi^2 + 
V(\ax,x_2), \nn \\
2 V(\ax,x_2) &=&  \frac{\cA^2}{X^2} + s x_2 + \la ( 3 x_2^2  - 2 \ax^4 ) 
\eea
It can be checked that the remaining Hamiltonian equations of motion
$\9 H / \9\ax = - \dot \pi $, $ \9 H / \9 x_2 = - \dot \pi_2 $ are in fact compatible with 
the original set of equations of motion.  
The corresponding effective Lagrangian is related to $H$ by a Legendre transformation, 
$L= \pi_2 \dot x_2 + \pi \dot \ax - H .$ Eq.\ (\ref{33a}) supplies the necessary relations 
of canonical and kinematical momenta, and we finally get 
\beq
L = \frac{1}{2}   \dot X ^2  +  \frac{1}{2}  \dot \ax^2  - V( \ax,x_2 ) .
\eeq
In terms of the coordinates $\ax,X$, the Lagrangian is composed 
of a kinematical kinetic contribution, a potential 
$(s X^2 + s\ax^2 + \la ( 3 X^4 + 6 X^2 \ax^2 + \ax^4 ) )/2 $ and  
an additional term $\cA^2 / (2X^2 )$ having the form of 
the energy of circular motion with angular momentum $\cA$. 

\subsection{The small coupling limit and parametric resonance.} 

In the strict limit $\la=0$ the equations of motion (\ref{29},\ref{30}) decouple trivially.
However, that limit does not provide a sensible starting point for a perturbative
expansion since the perturbation series turns out to be non-analytic around zero
coupling. We thus concentrate on the limit $\la \to 0^+$  
in which the equation of motion for $\ax$ has the structure 
\beq 
\ddot \ax +  \ax \om^2 (t)  =0  \lab{33}
\eeq 
of a linear oscillator with a time dependent frequency-factor. It can be seen by 
 completely eliminating $x_2$ from the system (\ref{29},\ref{30}) that 
for finite $\ax$
the quantities $x_2,p_2$ and $w$ consistently have to be taken of 
order $1/\la$. Consequently, also the energy being a function of 
those quadratic expectation values grows with the inverse coupling
and will be rescaled into the finite quantity $\ih = 18 \la h +1 $. 
The time dependent mass-term $\om^2 = ( s  + 2 z \sqrt{\ih}  ) /3 $ 
can be rewritten in terms of the function
$z = ( 9 \la x_2 + s ) /\sqrt{\ih}$ which is a solution of 
$z'' + z^2 - 1 = 0$,
where the prime denotes differentiation with respect to the rescaled
time $\tau = t ( 2 \sqrt{\ih})^{1/2}  $. Note that in this limit, there is no
back-reaction on the frequency-factor by the motion of the position 
coordinate. A first integral is given by
\beq
\frac{1}{2} (z')^2 = A  + z - \frac{z^3}{3} \lab{udot} 
\eeq
where $A$  encodes the initial conditions of $x_2 $. 
Solutions for $z(\ta)$ are periodic if $| A | < 2/3$ with 
minima in the strip $|z| <1$ and maxima for $1<z<2$.

\begin{figure}
\begin{minipage}{3in}
\resizebox{3in}{!}{\includegraphics{fig1.p.epsi}  }
\label{yplot}
\caption{Time evolution of the effective mass $\om^2(t)$ (dashed line) 
and $\ax(t)$ with parameters 
$\ih=3.1, \,z_0=-0.56, \ax_0=1,\, \ax'_0=z'_0=0 .$}
\end{minipage}
\hspace*{\fill}
\begin{minipage}{3in}
\resizebox{3in}{!}{\includegraphics{fig2.p.epsi}  }
\label{eex1}
\caption{The same as in Fig.\ 1, but with $\ih=1.84$ .} 
\end{minipage}
\end{figure}

For physical solutions of the system (\ref{33}--\ref{udot}) we have to 
keep in mind that only positive values of $x_2$ can be generated by
hermitian density matrices\footnote{$\deldel x > 0$ is trivially valid 
since $\la x_2$ is of the same order as  $\ax$.}.
Consequently, $z \sqrt{\ih} -s  $ 
has to remain positive in the course of time evolution. 
In the case of the unbroken potential $s =1$, 
this further restricts motion of $z$ to the strip $2 < 2 z \sqrt{\ih} < 
( -1 + \sqrt{6+3 \ih } ) $ where the bounds correspond
to the local extrema of a periodic motion during which 
$\om^2 $ remains positive. 

The more interesting case is the broken phase $s  =-1 $. There, the 
solutions which fulfill the positivity requirement for $x_2$ can 
now be found in the domain $-2 < 2 z \sqrt{\ih} < ( 1 + \sqrt{6 + 3 \ih } ) $
and correspond all to oscillatory $z$. However, if the initial 
conditions are varied, the behavior of $\om^2$ becomes quantitatively different.
In particular, in the range  $1 < 2 z \sqrt{\ih} < ( 1 + \sqrt{ 6 + 3 \ih ) }$, the 
factor $\om^2 $ is positive but becomes negative for 
$ -2 < 2 z \sqrt{\ih} < 1 .$ In the first region, the motion of 
$\ax$ is quasi-harmonic with variable frequency, but turns into exponential 
behavior if the lower strip of $z$ is reached by time evolution. Physically,
oscillatory and tunneling phases alternate. Negative square of frequency 
flips the sign of the curvature of  $\ax$
and bends trajectories in a direction opposite to the oscillations. 
However, since there is no back reaction on the motion of $\ax$ from the 
frequency-factor, the tunneling phases are short, and 
exponential growth stops before the numerical value of $\ax$
can become large. The period of $\ax$ is roughly 
determined by $\om^2$ averaged over one period of $z$.
Depending on the quotient of those two
periods, a variety of resonance phenomena
can appear. Figure (1) displays the case of a frequency coefficient 
close to $1:2$. The periodic kick of the motion of $\ax$ results 
in large resonant amplitudes. Figure (2) shows a
frequency quotient of $3:8$. The dents in the motion of 
$\ax$ correspond to tunneling phases.  

\newpage

\section{Second order Motion} 

Realistic thermodynamic behavior includes entropy
generating processes which necessitates the inclusion of the 
non-local integral contributions in (\ref{final}). The evaluation
of that term, however, requires an approximation for the 
evolution operator $T(t,t')$. A strict expansion in 
powers of the coupling constant to second order amounts to the 
replacement $L\to L_0$ in the definition of $T(t,t')$ which has been
applied recently in the context of field theory \cite{anas97}.
On account of the dynamical closure,  $T(t,t')$ even 
exponentiates to the product of the free unitary time 
evolution operator with $Q$.
That approximation, however, suffers from 
non-conservation of the energy (\ref{eint}) at secular scales. 
A numerical integration shows that the runaway of energy 
sets in even at times before the interesting 
transition from regular motion to the stochastic
entropy generating  phase appears. Moreover, since 
energy turns out to increase even 
exponentially, the expectation values also 
grow in an unphysical manner. A first attempt to cure
the situation is to replace the free Hamiltonian in 
$T(t,t')$ by a resummed quadratic one.  At first sight this
concept appears promising, since it still benefits from a
dynamic closure relation. We discuss this approach 
first before we turn to a further essential modification.  

\subsection{Resummed Hamiltonian approach} 

The resummed first order Hamiltonian is defined to be of at most
quadratic in position and momenta but with explicit time dependence
through coefficients $(a,\ldots,e)$, and has to reproduce the 
equations of motion ({\ref{eom1}--\ref{eom5}). We make the Ansatz 
$H^r (t) = H_0 + \la H_1^r (t) $ with 
\beq 
H^r_1(t)  = a p + b x  + c ( x p + p x ) + d x^2 + e p^2 .
\eeq
Plugging this expression into (\ref{final}) and keeping terms to first order
in the coupling, the equations of motion turn into 
\beq
\begin{array}{rclrcl}
\dot \ax &=&  \ap ( 1 + 2 \la e )  + 2 \la c \ax + \la a ,  & \hspace{-5em}
\dot \ap & = &- \ax ( s + 2 \la d ) - 2 \la  c \ap - \la b ,\nn \\ 
\dot x_2 & = &  w ( 1 + 2 \la e )  + 2 \la  a \ax + 4 \la c x_2 , & \hspace{-5em}
\dot  p_2 & = &  - w ( s + 2 \la d ) - 4 \la c p_2 - 2 \la b \ap , \\ 
\dot w  & = & 2 p_2 ( 1 + 2 \la e ) - 2 x_2 ( s + 2 \la d ) - 2 \la b \ax +  2 \la a \ap ,
\end{array}  \lab{sysr}
\eeq
which when compared with the original first order set of equations defines
\beq
a =  \xi ( 2 \ap x_2-w \ax ),\quad 
b =  - 4 \ax^3 + \xi ( 2 p_2 \ax -\ap w ) ,\quad 
c =  \frac{1}{2}\xi W^2  , \quad d =  3 x_2 - \xi P^2 , \quad 
e = -\xi X^2.
\eeq
The one-parameter solution parameterized by $\xi $ is a consequence of time 
independent uncertainty which still is an integral of motion of Eq.\ (\ref{sysr}). 
The remaining freedom 
has to be fixed by the condition that $H^r$ is a constant of time which
is equivalent to the fact that the expectation value of $H^r$ equals 
the expectation value of $H$,
\beq 
\xi = \frac{ 3 ( 2 \ax^4 - x_2^2 ) }{ w^2 -4 x_2 p_2 + 6 ( x_2 \ap^2 +
\ax^2 p_2 -  w \ax \ap  )   } . 
\eeq

Resummation is achieved by absorbing the effective Hamiltonian 
into the new free one, and compensating by an interaction counter term,
\beq
H = H^r + H_2,  \quad H_2 = H_1 - \la  H_1^r (t). 
\eeq
In the integrals of the dissipative contributions in (\ref{lls}),
we have to replace $L_1$ by $L_2$.
However, the set of observables
$\cE$ is still dynamically  closed with respect to $H^r$ such that 
only $H_1$ of $H_2$ contributes. 
Finally, the resummation enters in the expansion of the 
evolution operator $T(t,t')$ where we replace $L \to L^r(t) $ in 
(\ref{Tgl}). We have to consider the particular combination  
$T(t,t') \circ Q(t')$ appearing in the entropy generating 
contribution which can be integrated formally directly from its definition, 
\beq
T^r(t,t') \circ Q(t')  = 
\cT \exp \left( - i \int_{t'}^t d\ta Q(\ta) \circ  L^r(\ta) \circ  {} \right) Q(t') .
\eeq
At that point, dynamical closure guarantees the important relation 
$Q(t'') \circ L^r (t) \circ Q(t') =  L^r (t) \circ Q(t')$ valid even in the 
case where the time arguments of the projectors differ. 
All $Q$ but the one at the most r.h.s in the summands of the formal
exponential can be dropped, and we arrive at the desired relation 
\beq T^r(t,t') \circ Q(t')  =  \cU(t,t') \circ  Q(t'), \quad 
\cU(t,t') = \cT \exp \left( - i \int_{t'}^t d\ta L^r(\ta) \circ {} \right) .
\eeq
We want to point out  that this relation does not automatically
imply the equality of $T^r(t,t') $ and $\cU(t,t')$. The replacement  
is non-trivial also in the sense that acting on the projector,
the action of a non-unitary operator is replaced a unitary one. 

It turns out to be useful to let $\cU$ act on the operators to 
its left,
$ \tr  \left( (  \cU^\dag (t,t')\circ   \cO_1 ) \circ  \cO_2  \right) = 
 \tr \left( \cO_1 \circ \cU(t,t')\circ  \cO_2 \right) $.
On account of the two properties of $L^r$ being a derivation and anti-hermitian,
the adjoint action simply amounts to exchange time arguments,
$ \cU^\dag (t,t') = \cU(t',t) $. We introduce the evolved operator
$ \cO (t,t') =  \cU^\dag (t,t') \circ \cO  $
which solves 
\beq
\9_{t'} \cO (t,t') = - i L^r (t') \circ \cO (t,t') \lab{eomO} \eeq
and is related to the Dirac representation induced from $L^r$ by 
$\cO_D (t,t') = \cU(t,t') \circ  \cO = \cO(t',t) .$
Since the Liouvillian is a derivation, it suffices to investigate on the 
evolution of $x,p,\1 $ representing an operator basis. Dynamical
closure guarantees the success of the Ansatz
\beq
x(t,t') = x \a_x(t,t')  + p \b_x(t,t')  + \g_x(t,t'), \quad
p(t,t') = x \a_p(t,t')  + p \b_p(t,t')  + \g_p(t,t')
\lab{heis}
\eeq
with $\a_x(t,t'), \ldots,\g_p(t,t') $ to be determined by coefficient comparison.
That gives rise to the system
\bea
\9_{t'}  \a_x(t,t') &=& -2 \la  c(t')  \a_x(t,t') +  ( s  + 2 \la d(t') ) \b_x(t,t') \nn \\
\9_{t'}  \b_x(t,t') &=& - ( 1 + 2 \la e(t') )  \a_x(t,t') + 2 \la c(t')  \b_x(t,t') \nn \\
\9_{t'}  \g_x(t,t') & = & - \la a(t') \a_x(t,t') + \la b(t')   \b_x(t,t') . \lab{dsalbe}
\eea
Integrals are found to read 
\beq
\a_x(t,t') \ax(t') + \b_x(t,t') \ap(t') + \g_x(t,t') = \ax(t) , \lab{59}
\eeq
which corresponds to the quantity $\tr ( x(t,t') \cR(t') ) $ independent of $t'$, and 
\beq
\a_x^2 (t,t') X^2(t') + \b_x^2 (t,t') P^2(t') + \a_x (t,t') \b_x (t,t') W^2(t') 
= X^2(t)  , \lab{60} 
\eeq
expressing the $t'$ independence of $\tr ( x(t,t')  x(t,t') \cR(t') ) $. We remark that 
these constants imply unitary evolution of the accompanying operator,
$\cU(t,t') \cR(t') = \cR(t) $, and are compatible with a time independent uncertainty
product. The complete solution of (\ref{dsalbe}) can be parameterized by the angle variable
\beq
\Psi(t,t') = \int_{t'}^t d\ta \cA \left( X^{-2} (\ta) - 2 \la \xi(\ta)\right). \lab{44}
\eeq
We find 
\beq
\a_x(t,t')  = \cA^{-1} X(t) P(t')  \cos( \Psi (t,t')+ \eta(t') ) , \quad 
\b_x(t,t')  =    \cA^{-1} X(t) X(t') \sin( \Psi(t,t') )  \lab{45} 
\eeq
with 
\beq
\cos \eta(t) =  \cA X^{-1} (t) P^{-1} (t)   
, \quad  \sin \eta(t) = W^2(t) X^{-1} (t) P^{-1} (t)/2 , \lab{444}
\eeq
which satisfies the initial conditions $\a_x(t,t)=1 ,\, \b_x(t,t) =0$. The coefficients
for $p(t,t')$ are also solutions of the differential equations (\ref{dsalbe}), but with 
integration constants
$\ap(t)$ and $P^2(t)$ at the r.h.s.\ of (\ref{59},\ref{60}). The corresponding 
angle variable gets shifted to $\Psi(t,t') - \eta(t) + \pi/2 $ and we 
get
\bea
\a_p(t,t') &= & -\cA^{-1} P(t) P(t') \sin( \Psi(t,t') - \eta(t) + \eta(t')  ), \nn \\
\b_p(t,t') & = & \cA^{-1}  P(t) X(t') \cos( \Psi(t,t')  - \eta(t) )  \lab{46} .
\eea

Now all necessary tools are at hand to evaluate the integrals in 
(\ref{lls}) and (\ref{eint}) necessary solve the integro-differential equation of motion
in a resummed Hamiltonian approximation. We have integrated the system 
numerically and still find non-conservation of energy at typical time scales of order
$ 2 \pi/(4 \la) $ for $\la \ll 1$. The factor $2 \pi$ corresponds to the natural period
of the oscillator with unit circular frequency, and  $1/4$ can be identified to coming from 
the fourth powers of the harmonic functions in $\a,\b$
due to the quartic interaction Hamiltonian. 
At a technical level, the problem is that quadratic
Hamiltonians, even with time dependent coefficients, automatically 
conserve the uncertainty $\cA^2$, and motion stays within this equivalence 
class fixed by the initial conditions. Plugging that approximation into the 
entropy generating integrals of the equations of motion, 
the dissipative corrections, although formally of 
second order, do not remain small at large time scales.
The effective unitary evolution operator generated by an effective
quadratic Hamiltonian cannot be used as a sensible approximation
in the entropy generating corrections. The inconsistency becomes apparent
when one compares the uncertainty $\cA$ determined by the initial conditions
with the actual uncertainty expressed by the evolved value of the 
observables.

\section{Consistent effective approximation.}

Although the effective Hamiltonian constructed in the previous 
section turns out to be inconsistent with 
time--variant uncertainty and time evolution at large time scales, 
it still approximates time evolution locally in time
at short time scales. It can be seen from numerical integration
that the quadratic Hamiltonian adjusted to a set of 
given quantities $x(t),\ldots,w(t)$ considered as initial 
conditions evolves them quite accurately.
Local in time, motion appears to be generated by a quadratic
Hamiltonian which can be used to approximate only the enveloping 
curves of the true trajectories, but the locally best approximating 
effective Hamiltonian does not evolve to the
enveloping Hamiltonian being the best approximation at a later time.
One may argue that a non-quadratic extension may solve the 
problem of energy conservation. However, on account of the 
basis of observables chosen, the accompanying density 
operator still remains quadratic, and the exact Wick theorem valid,
such that the non-quadratic extension effectively boils down to 
the quadratic effective theory. 

We thus abandon the query for an effective Hamiltonian completely 
in favor of the construction of an effective time evolution operator.
Physically, we have to admit effective equations of motions for the 
observables which are not generated by a Hamiltonian function.
This step is also necessary to account for irreversibility.

The complete effective time evolution operator $\cV (t,t')$ be 
subject to the following assumptions. \\
(a)
Transitivity, $\cV(t,t') \circ  \cV (t',t'') = \cV (t,t'')$ \\
\noindent
(b) Initial value,  $\cV (t,t) = \1 $. \\
\noindent
(c) Its adjoint acts like an exponentiated derivative operator on products, i.e.
$\cV (t,t')^\dag  \circ  ( \cO_1 \cO_2 ) = ( \cV^\dag (t,t')\circ \cO_1 )  (\cV^\dag (t,t')\circ \cO_2 )$.
Time evolution commutes with forming operator products. 

We will further specify to the 'collisionless'  approximation of $\cV(t,t')$ by 
the additional assumptions that \\ 
\noindent
(i) --- the operator linearly maps the basis $x,p,\1$ onto itself. 
That replaces the notion of 'free motion', and generalizes dynamical closure 
to non-Hamiltonian evolution. \\
\noindent
(ii) --- $\cV(t,t')$ is compatible with time evolution of the accompanying
density operator, i.e.\ $\cV(t,t') \circ \cR(t') = \cR(t)$. That condition 
actually defines $\cV$ consistently with the equations of motion, and the 
evolution of $\cR$, which is just 
contrary to the strategy of the previous chapter, where
the evolution operator generated by the effective Hamiltonian 
was responsible for time evolution of $\cR$. 

At a technical level, time dependent 
uncertainty $\cA(t)$ is introduced into the coefficients $\a,\b$ which 
are no longer subject to the set of differential equations (\ref{dsalbe}). 
If we define the evolved operators by $\cO(t,t') = \cV^\dag (t,t') \circ \cO$, the basis
still can be expanded as in (\ref{heis}). The requirements (ii,c) automatically
lead to the relations
\beq
\begin{array}{rclrcl}
\tr ( x(t,t') \cR(t') ) &=& \ax(t), &  \tr ( x(t,t') x(t,t')  \cR(t') ) &=& x_2(t), \nn \\
\tr ( p(t,t') \cR(t') ) &=& \ap(t), &  \tr ( p(t,t') p(t,t')  \cR(t') ) &=& p_2(t), 
\end{array}
\eeq 
which we identified as constants of motion of the resummed evolution
equations. In addition to these, we find 
\beq 
\tr ( ( x(t,t') p(t,t')+p(t,t') x(t,t') )  \cR(t') ) = w(t) 
\eeq
the r.h.s.\ of which was fixed by initial conditions in resummed Hamiltonian 
dynamics. These equations are sufficient to express 
the evolution coefficients of $x$ and $p$ in a manner analogously to 
(\ref{45},\ref{46}) together with (\ref{444}), but with $\cA$ replaced by the 
actual uncertainty $\cA(t')$ in (\ref{45},\ref{46}) and $\cA(t)$ 
in (\ref{444}). 
The angle variable $\Psi(t,t')$, however, remains undetermined for the moment. 

In the evaluation of $T(t,t') \circ Q(t')$, we consistently have to 
replace the resummed unitary operator $\cU(t,t')$ by $\cV(t,t')$. 
The dissipative contributions $\Si_\nu (t) $ and $\ep(t) $ 
are now found to read ($\Psi \equiv \Psi(t,t'), \, \ax \equiv \ax(t),
\, \ax' \equiv \ax(t'),\ldots $) 
\beq
\Si_\nu(t) = \la^2 \int_0^t dt' \s_\nu(t,t') , \qquad \ep(t) = \la^2 \int_0^t dt' \ep(t,t') ,
\eeq
with
\bea
\s_p(t,t') & = & 
 6 \cA'^{-3} X^3  X'^3 \sin \Psi 
     \left( 12 \cA'^2 \cos^2  \Psi - \sin^2 \Psi \right) \ax' ,
\nn  \\
\s_{p_2}(t,t') &= & 12  \cA'^{-3} X'^3  X^2 P  \times \nn \\
  && \hspace{-2em}  \left( X X' \left( 
        4 \cA'^2  ( 2 \cos(2 \Psi - \eta ) - \cos\eta ) \cos^2 \Psi -
         ( 2 \cos(2 \Psi - \eta ) + \cos\eta )\sin^2 \Psi \right)  \right. 
\nn \\ 
         &&   \hspace{-2em} \left. \,\, {}+  \ax \ax' \left( 6 \cA'^2 
( 3 \cos(2 \Psi - \eta)- \cos\eta )  
           \cos\Psi - 3 \cos(\Psi -\eta ) \sin^2 \Psi \right)  \right)  
\nn \\ 
         &&   \hspace{-2em}   \,\, {}+ 2 \ap \s_p(t,t') ,
\nn \\
\s_w(t,t')  & = & 48 \cA'^{-3} X^3 X'^3 \sin \Psi  \times  \nn \\ 
     && \, \left( X X' \left( 4 \cA'^2 \cos^2 \Psi -   \sin^2  \Psi  \right)  \cos \Psi  + 
               \ax \ax'  \left( 12 \cA'^2 \cos^2 \Psi   -   \sin^2 \Psi \right) \right)  ,
\nn \\
\ep (t,t') &= &  -\s_w(t,t') /8  .
\eea
The last equality is due to the particular form of the operator 
$x p+px $ which, when commuted with a homogeneous 
operator, evaluates to twice the degree of homogeneity in $x$. 

A complete solution involves also to determine the  unknown angle 
variable $\Psi(t,t') $.  Analogously to the construction of 
the effective Hamiltonian, we exploit the remaining freedom to 
require the energy to be a conserved quantity   
within our approximation scheme. Differentiating the 
corresponding expressions (\ref{eint}) with respect to time, on finds 
that the term $d\ep(t)/dt $ has to be compensated by the integrand of 
the term $\Si_{p^2}(t) /2 $. That gives rise to a differential equation for
$\Psi$ with solution
\beq
\Psi(t,t') = \int_{t'}^t d\ta \cA(\ta) X^{-2} (\ta)  .
\eeq  
It is a non-trivial result that the angle parameter can be expressed in
terms of an integral involving the history of $\cA$ and 
$X$. That nevertheless includes all contributions which can
be resummed in an effective evolution operator satisfying conditions
(i) and (ii) in a self consistent way.
Note also that the first order correction of (\ref{44}) can 
be absorbed into the time dependent uncertainty completely. Moreover,
since uncertainty is directly related to entropy (see Appendix), the approximation
consistently accounts for entropy variations but by construction keeps the 
energy constant in time. 

The complete set of integro-differential 
equations can now be evaluated numerically. A typical
time evolution is displayed in Figs.\ (3-6) for the initial conditions 
$\ax(0) = 1,\,  x_2(0)= p_2(0)=2 , \, \ap(0) =w(0)=0$ with $\la = 0.1 $ 
and symmetric potential $s=1$.

\begin{figure}[t]
\begin{minipage}{3in}
\resizebox{3in}{!}{\includegraphics{vlp.dx.epsi}  }
\caption{Time evolution of the uncertainty of the
position operator $X(t)$.}
\end{minipage}
\hspace*{\fill}
\begin{minipage}{3in}
\resizebox{3in}{!}{\includegraphics{vlp.dp.epsi}  }
\caption{Time evolution of the uncertainty of the
momentum operator $P(t)$.}
\end{minipage}
\end{figure}

\begin{figure}
\begin{minipage}{3in}
\resizebox{3in}{!}{\includegraphics{vlp.x.epsi}  }
\caption{Time evolution of the position average $\ax(t)$. }
\end{minipage}
\hspace*{\fill}
\begin{minipage}{3in}
\resizebox{3in}{!}{\includegraphics{vlp.S.epsi}  }
\caption{Time evolution of the relevant entropy $S(t) $. }
\end{minipage}
\end{figure}

It can be proven in the general setting that motion 
is such that the entropy is always larger or equal than its initial value.
We find that our approximation respects that property, but 
fluctuations at intermediate times are possible. 
There is no sign that the system approaches a static point at large time 
scales which is not to be expected in general.  

The numerical strategy to solve the consistency problem for the
integro--differential equation was the following. Suppose we knew
the exact set of variables in the time range $(0,t)$. Then we benefited 
from the enveloping Gaussian dynamics and  
predicted the evolution by integrating the first order 
system with the initial data at $t$ for a time interval $(t,t+\delta t)$,
where $\delta t$ is typically of the order of a period of $\ax$.

That prediction can be improved iteratively for $(t,t+\delta t)$
by plugging it into the integral at the r.h.s.\ which 
represents an inhomogeneity for the differential equation. 
Note that the integral
expands over the complete history of the observables
which poses a problem in numerical integration over large
time scales. 
Luckily, the integrand can be split into a sum of products
of functions depending solely on $t$ or $t'$, such that 
the integrals can be reused in the extension procedure of time 
range. For times including a large number of 
quasi--periods of the variables, the contribution of the integral, 
although evaluated exactly, increasingly exhibits stochastic 
behavior. 
The method proposed thus supplies a tool to exactly 
calculate quasi--stochastic dissipative behavior which 
can be used to test a priori assumptions about stochastic forces.

\section{Conclusion and Outlook}
We proposed a method to calculate the time evolution of 
observables beyond the effective Hamiltonian approximation in
a self--consistent systematic way thus accounting for real 
dissipative processes. An application of this approximation scheme 
to field theories is planned. In particular, we are interested in 
the influence of real collisional processes on long term evolution,
and expect to be able to decide whether dephasing or proper 
entropy generating processes dominate in infinite--dimensional
systems. A generalization to gauge and fermion  systems is planned,
and we intend to develop a systematic expansion to higher orders.
Applications may also include dynamical problems in cosmology 
such as inflationary phases.

\begin{appendix} 

\section{Entropy, Uncertainty and Effective Lagrange Multipliers}

We relate the Lagrange multipliers $\mu_\nu$ and the relevant 
entropy $S=\mu_\nu g_\nu$ to
the set of observables $\ax,\ap,x_2,p_2,w$. One considers 
\beq
[\cF_\mu ,\cR] = -\mu_\s \int_0^1 dx \cR^x  [ \cF_\mu , \cF_\s] \cR^{1-x } , 
\eeq
which has vanishing trace.  That gives rise to the set of equations
$\mu_\s g_\rh  \G^{\mu \s }_\rh =0 , \, [\cF_\mu , \cF_\s ] =    \G^{\mu\s }_\rh \cF_\rh $
which are not independent, but lead to 
\beq
\mu_x = \eta ( \ax P^2  - \frac{1}{2} \ap W^2  ), \quad 
\mu_p = \eta ( \ap X^2 - \frac{1}{2}  \ax W^2  ), \quad 
\mu_{x^2} = - \eta P^2 /2 , \quad \mu_{p^2} = - \eta X^2 /2 \lab{gamm}
\eeq
where $\eta = 4 \mu_w / W^2 $. Thus, one more relation is called for to 
determine the remaining unknown.
To that end, we consider the eigenvectors $\cG_\pm $ of $\log \cR$ being linear 
combinations of the operators $\cF_\nu$. The eigenvalue equation  
$[ \cG_\pm , \log \cR ] = \xi_\pm \cG_\pm $ implies  
$ \tr ( \cG_\mp  \cG_\pm \cR ) = \e ^{\xi_\pm} \tr ( \cG_\pm \cG_\mp \cR ) $,
with the conjugate pairs of eigenvectors one of which  is found to have the eigenvalues
$\xi_\pm = \pm 2 \sqrt{\mu_{x^2} \mu_{p^2 } - \mu_{w}^2 } = \pm \eta \cA $.
On the other hand, the traces can be evaluated using Wick's theorem, and one finds 
\beq
\log \left( \frac{2 \cA - 1 }{   2 \cA + 1   } \right) = \eta \cA ,
\eeq
which expresses the remaining unknown $\eta$ in terms of the observables. 
From the constant normalization $\tr \dot \cR(t)=0 $ 
we get $\dot \mu_0 + \dot \mu_i g_i =0 $, and thus $\dot S = \mu_i \dot  g_i $.
The summands can be combined in terms of $\cA(t)$ and we finally 
get 
\beq
S(t) = ( \cA(t)  + \frac{1}{2} ) \log ( \cA(t)  + \frac{1}{2} ) -
           ( \cA(t)  - \frac{1}{2} ) \log ( \cA(t)  - \frac{1}{2} )
\eeq
where  a possible integration constant was chosen such that entropy vanishes
at the minimal possible uncertainty corresponding to $\cA=\frac{1}{2} $.  $S(t)$ is 
a monotonically increasing function of $\cA(t)$. This relation was found in a different 
context earlier \cite{anas94}.

\end{appendix}

\end{document}